\documentclass[sigconf, nonacm]{acmart}
\setcopyright{none}
\AtBeginDocument{%
  }

\begin{document}

\title[Towards Reliable AI-Assisted Analog Design]{Towards Reliable AI-Assisted Analog Design: Template-Constrained LLM Agents for SAR ADC Generation}

\author{Dimple Vijay Kochar}
\affiliation{%
  \institution{Massachusetts Institute of Technology}
  \city{Cambridge, MA}
  \country{USA}}

\author{Hae-Seung Lee*}
\affiliation{%
 \institution{Massachusetts Institute of Technology}
 \city{Cambridge, MA}
 \country{USA}
}

\author{Anantha P. Chandrakasan*}
\affiliation{%
  \institution{Massachusetts Institute of Technology}
  \city{Cambridge, MA}
  \country{USA}
}

\renewcommand{\shortauthors}{Kochar et al.}

\begin{abstract}
While Large Language Models (LLMs) have demonstrated significant capability in software code generation, their application to analog Electronic Design Automation (EDA) is bottlenecked.
Owing to limited circuit topology understanding and data, directly prompting LLMs and multimodal models leads to hallucinations and failure to produce schematics capable of passing rigorous SPICE simulations, as we show in our work.
Instead, we propose an end-to-end, multi-step LLM agentic framework \modelName, capable of generating a functional Successive Approximation Register (SAR) Analog-to-Digital Converter (ADC) that successfully passes simulation validation.
To adhere to the rigid constraints of analog design, we utilize expert knowledge to ground the LLM in its planning, selection, parameterization, and iterative modification.
As part of \modelName, we introduce Template-Constrained Generation - which unlike other template-based works - builds towards a more generalized SAR ADC generation flow.
We demonstrate a strong proof-of-concept of our framework by developing SAR ADCs across technology nodes and input specs.
Overall, our expert-knowledge grounded multi-step agentic \modelName{} establishes a pragmatic foundation for integrating LLMs into reliable analog design methodologies.
\end{abstract}


\newcommand{\SideNote}[2]{\todo[color=#1,size=\small]{#2}} 
\newcommand{\tanmay}[1]{\SideNote{orange!40}{#1 -> tanmay}}
\newcommand{\dimple}[1]{\SideNote{purple!40}{#1 -> dimple}}

\newcommand{\modelName}{\textsc{ATLAS}}

\maketitle

\section{Introduction}

Continuous scaling of technology nodes and ever-increasing demand for mixed-signal Integrated Circuits (ICs) necessitate efficient, scalable design methodologies.
Within these, Analog-to-Digital Converters (ADCs) - particularly Successive Approximation Register (SAR) ADCs - are critical components that bridge the physical and digital domains.
However, while digital Electronic Design Automation (EDA) workflows have achieved high levels of abstraction and automation \cite{ref2}, analog design remains largely manual, iterative, and heavily intuition-driven \cite{ref1}.
This lack of scalable automated synthesis for analog circuits creates a severe bottleneck in modern System-on-Chip development.

\begin{figure}
    \centering
    \includegraphics[width=0.98\linewidth]{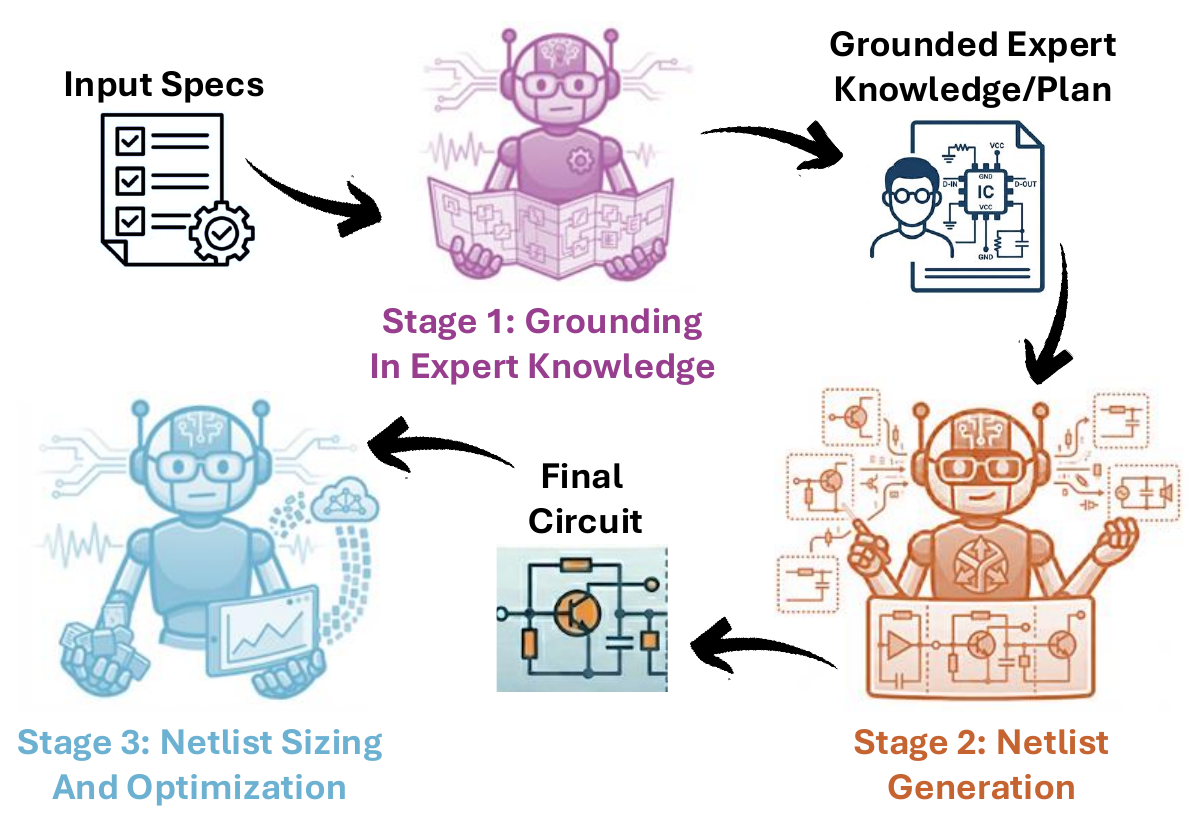}
    \Description{High-level workflow of our proposed multi-step LLM-based agentic framework \modelName{} for creating SAR ADCs from user-defined input specifications in an end-to-end automated manner.}
    \caption{High-level workflow of our proposed multi-step LLM-based agentic framework \modelName{} for creating SAR ADCs from user-defined input specifications in an end-to-end automated manner.\protect\footnotemark}
    \label{fig:teaser}
\end{figure}

\footnotetext{We acknowledge the utilization of Generative AI - specifically Nano Banana - for creating illustrations and figures in our paper.}

Recently, Large Language Models (LLMs) have demonstrated transformative capabilities in automated software engineering, logic synthesis, and hardware description language (HDL) generation and verification \cite{swe, ref2, chipnemo, grposmu}.
However, its utility for analog EDA has been restrictive, owing to the rigid, continuous-domain physical constraints in analog EDA that differ fundamentally from discrete digital logic or software code.
To this end, we conduct a deep qualitative study to explore the utility of existing LLMs/multimodal models for designing analog schematics from scratch.
Our study demonstrates that directly prompting LLMs in an unconstrained manner frequently results in hallucinations, producing physically unviable schematics with connectivity errors.

\begin{figure*}[t]
    \centering
    \includegraphics[width=0.9\linewidth]{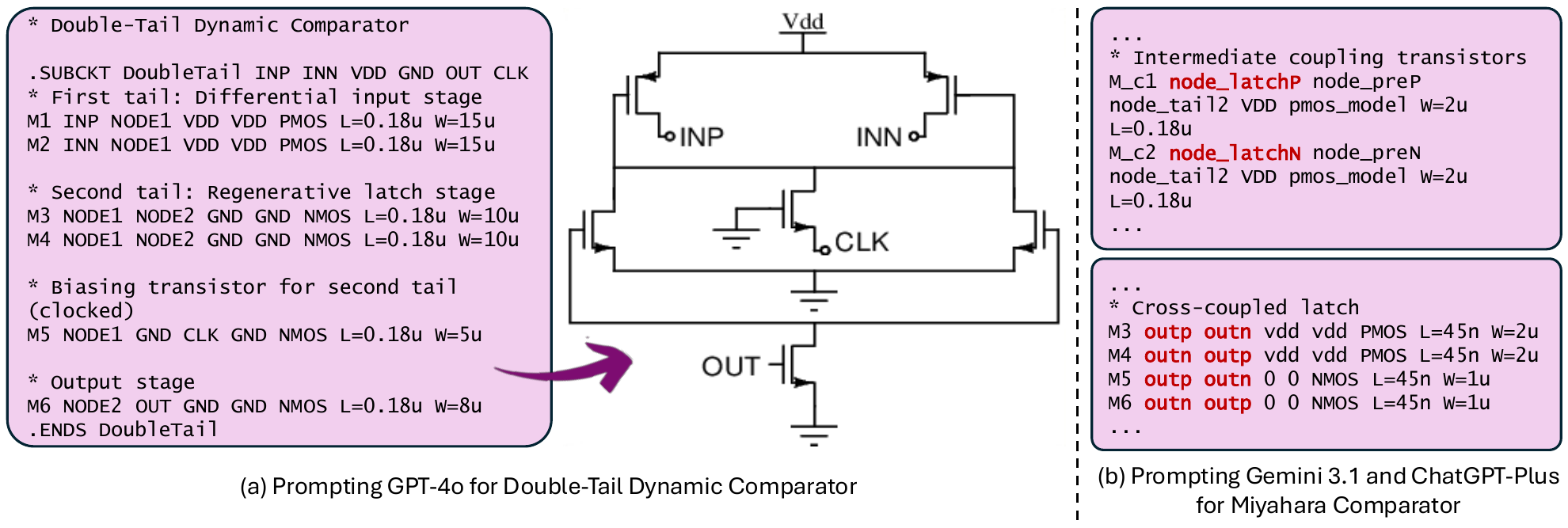}
    \Description{Highlighting several limitations of directly prompting LLMs for netlist generation. On the left (a), we highlight several severe issues like shorted gates, input to drains, etc. in the GPT-4o created netlist. On the right (b), we highlight issues like missing connections and missing intermediate cross-coupling when prompting frontier LLMs.}
    \caption{Highlighting several limitations of directly prompting LLMs for netlist generation. On the left (a), we highlight several severe issues like shorted gates, input to drains, etc. in the GPT-4o created netlist. On the right (b), we highlight issues like missing connections and missing intermediate cross-coupling when prompting frontier LLMs.}
    \label{fig:prompting-llms}
\end{figure*}

To address these critical limitations, this work shifts the paradigm from unconstrained circuit generation to grounded, feedback-based synthesis.
Rather than forcing the LLM to generate complex analog topologies from scratch, we propose an end-to-end, multi-step LLM agentic framework \modelName{} - \textbf{A}gentic \textbf{T}emplate-constrained \textbf{L}LM-based \textbf{A}DC \textbf{S}ynthesizer (illustrated in Figure~\ref{fig:teaser}).
The core distinction of our framework lies in the grounding with expert knowledge across various components for planning, selection, stitching, re-configuration, testbench generation, etc., designed specifically for reliable analog circuit synthesis.
We also introduce a generalized Template-Constrained generation that enables \modelName{} to select component-wise expert templates and structurally modify them to meet the target specifications.
Feedback from our rule-based verification and simulation engines allows \modelName{} to self-debug and improve the circuit.
Overall, by constraining the generative search space, our agentic framework prevents the LLM from diverging into physically impossible configurations while still harnessing its capacity for rapid iteration and code generation.

To demonstrate the efficacy of our framework, we utilize it to synthesize a low-power 8-bit SAR ADC on Cadence GPDK-45nm technology node.
Our agentic framework generated ADC successfully meets the input specs with an ENOB of 7.59.
We further show \modelName's generalizability by generating three additional SAR ADCs:
(1) the original 8-bit SAR ADC transferred to TSMC 65nm (generalization across technology nodes),
(2) a 4-bit SAR ADC (generalization across input specifications), and
(3) a 10-bit SAR ADC (capability of bit modification).

In conclusion, our work makes the following contributions:
(1) we qualitatively showcase the limitations of direct prompting of LLMs and multimodal models for analog circuit generation,
(2) we propose a multi-step LLM agentic framework \modelName{} to reliably ground the circuit generation, and
(3) we demonstrate the efficacy of our framework by synthesizing four SAR ADCs across technology nodes and input specs.
Overall, our work serves as a strong proof-of-concept for reliable LLM-assisted analog design.

The remainder of the paper is organized as follows.
Section~\ref{sec:related-works} discusses the past works in this direction.
Section~\ref{sec:zero-shot-failure} shows various qualitative results highlighting the limitations of directly prompting LLMs for ADC netlist generation.
Section~\ref{sec:methodology} provides a detailed description of our proposed multi-step LLM agentic framework \modelName.
Section~\ref{sec:main-synthesis} demonstrates the efficacy of our framework to generate a low-power 8-bit SAR ADC end-to-end, while Section~\ref{sec:generalization} highlights the generalizability of our work.
We conclude and discuss future work in Section~\ref{sec:conclusion}.

\section{Related Works}
\label{sec:related-works}

Here, we briefly discuss various works in the domain of automating analog design and utilizing LLMs/AI in circuit design.

\paragraph{Computer-Aided Design (CAD) for Analog Circuits}
Recent research has been dedicated to automating analog circuit sizing, with methodologies spanning from analytical equation derivations to Bayesian Optimization (BO) \cite{mandal2002cmos, zhang2020efficient}. 
Recently, there has also been a surge of works utilizing reinforcement learning (RL) methods \cite{settaluri2020autockt, wang2020gcnrl}.
Within the specific domain of Successive Approximation Register (SAR) ADCs, automation efforts have largely bifurcated into topology generation frameworks and algorithmic sizing. 
Generation frameworks frequently rely on template-based compilers \cite{opensar1, opensar2, ding2018hybrid} that use pre-defined topologies to automate layout assembly.
Alternatively, macro-based SAR ADC generation leverages digital standard cells \cite{seo2018reusable} to create synthesizable analog building blocks. 
On the algorithmic sizing front, recent optimization techniques explicitly targeting SAR ADCs include multi-agent reinforcement learning \cite{marl}, global-local optimization \cite{sizing1}, and automated sizing via analytical equations \cite{sizing2}.
Despite achieving high optimization efficiency, these approaches inherently function as sizing algorithms for a single, fixed SAR ADC, lacking the flexibility to easily generalize across varied architectural design spaces.

\paragraph{LLMs for Circuit Design}
Several works \cite{ledro, heart, eesizer} have utilized LLMs plugged with external optimizers for analog circuits sizing.
Specific to analog circuit design, several works \cite{lamagic, atelier} have demonstrated that general-purpose LLMs can not design even simple analog circuits, as we also concur through our case study of designing simple ADC components in Section~\ref{sec:zero-shot-failure}.
AnalogXpert \cite{analogxpert} utilizes in-context examples for topology utilizing sub-circuit connections, but demonstrates poor LLM performance.
Owing to lack of strong analog principles in mainstream LLMs, works like Artisan \cite{artisan} propose niche task-specific fine-tuning of LLMs to boost performance on simple and specialized circuits.
In a similar direction, ChipNeMo \cite{chipnemo} and AnalogSeeker \cite{analogseeker} propose data curation and train foundation models for digital and analog design, respectively.
In other directions, AnalogCoder \cite{analogcoder}, AmpAgent \cite{ampagent} and LADAC \cite{ladac} develop agentic LLM frameworks for schematic design of simpler circuits like amplifiers, oscillators, etc.
Majorly, most of these works have explored LLMs for simpler analog circuits, while we focus our work on designing SAR ADCs.

\section{Direct LLM Prompting Limitations}
\label{sec:zero-shot-failure}

In this section, we evaluate the capability of LLMs and multimodal LLMs to understand circuit specifications and create netlists for analog design when prompted directly in an end-to-end fashion.
By highlighting the limitations of existing LLMs, we motivate our agentic framework in Section~\ref{sec:methodology}.
Compared to other works \cite{heart, ledro}, we note that our work is not an exhaustive benchmark/evaluation of LLMs, but rather provides qualitative case studies to highlight some limitations of direct prompting.

\begin{figure}
    \centering
    \includegraphics[width=0.9\linewidth]{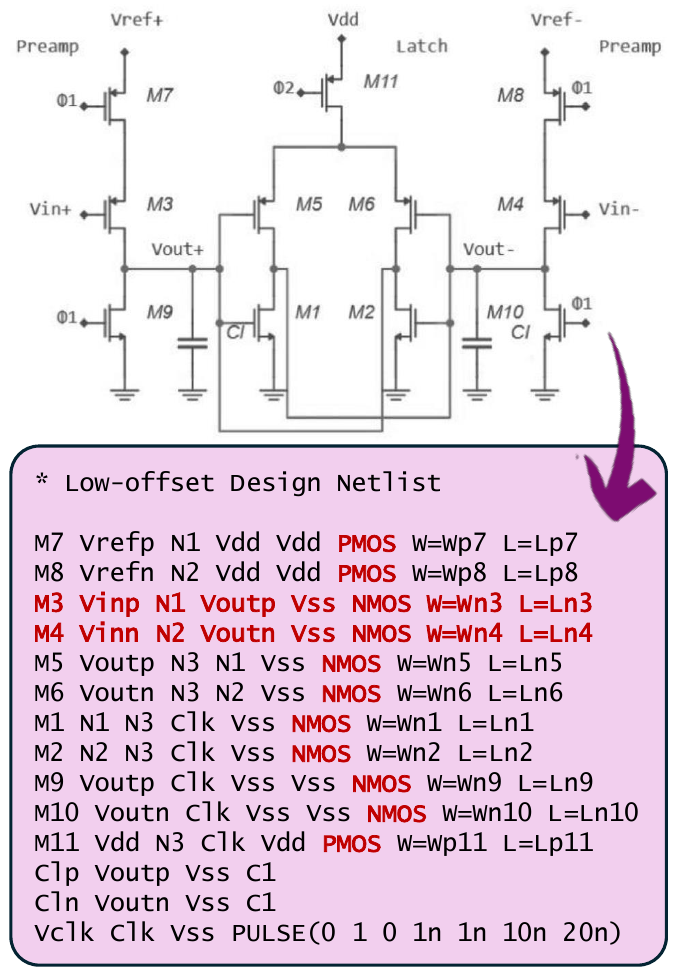}
    \Description{Highlighting the limitations of utilizing multimodal LLMs for netlist generation. In red, we highlight the wrong connections and incorrectly identify transistors.}
    \caption{Highlighting the limitations of utilizing multimodal LLMs for netlist generation. In red, we highlight the wrong connections and incorrectly identify transistors.}
    \label{fig:prompting-multimodal}
\end{figure}

\subsection{Prompting LLMs}

Synthesizing a SAR ADC from scratch is a formidable challenge even for human experts, making end-to-end generation by an LLM highly impractical.
Instead, as part of this case study, we prompt several LLMs to generate netlists for simpler components - specifically comparator - of the SAR ADC.
We present qualitative results from the LLM outputs in Figure~\ref{fig:prompting-llms}, highlighting errors and hallucinations.

First, we prompt GPT-4o \cite{gpt4o} to generate a double-tail dynamic comparator \cite{dtcomp}.
As shown in Figure~\ref{fig:prompting-llms} (a), there are multitudes of errors in the created netlist.
Some of them include:
(1) The gates of the input transistors are tied together at NODE1, completely negating any differential sensing capability, 
(2) The differential input signals (INP and INN) are incorrectly connected to the drain terminals of M1 and M2,
(3) This netlist contains its OUT at the gate of a transistor,
(4) The gate of the "clocked" biasing transistor (M5) is permanently tied to GND. Under standard conditions, this NMOS transistor will never turn on.
Inspection of the post-generation rationale reveals a severe lack of analog design knowledge, leading to hallucinations and incorrect assumptions, in turn causing the fundamental errors above.
We observe similar behaviors with same-sized LLMs from DeepSeek \cite{deepseek} and Gemini \cite{gemini}.

Under a higher compute budget, we also survey frontier models like Gemini 3.1 and ChatGPT Plus.
These models are much better at analog design fundamentals and avoid the basic mistakes made by the smaller models.
However, they also exhibit various errors when asked to generate the simpler comparator from Miyahara \cite{comp1}, as shown in Figure~\ref{fig:prompting-llms} (b).
For Gemini-Pro (top in (b)), we note that the generated netlist has nodes (node\_latchP and node\_latchN) that are not connected.
On the other hand, ChatGPT Plus (bottom in (b)) misses creating intermediate nodes and merges them to output nodes creating incorrect circuit logic.
We attribute some of these errors to failure in reasoning and reduced confidence, leading to hallucinations and incorrect connections.
Overall, we note that LLMs are getting better at understanding analog circuits and can be utilized for simpler circuits; their utility through direct prompting for creating SAR ADCs remains limited.

\subsection{Prompting Multimodal LLM}

Recently, LLMs have improved on various image understanding tasks and reasoning tasks \cite{yin2024survey}.
\cite{masala-chai} have explored the use of multimodal models for reasoning over simpler circuit images.
To this end, we conduct a small case study to study the capability of multimodal LLMs (specifically GPT-4o) to generate netlists from images of a low-offset comparator circuit from \cite{low-offset}.
We attempted direct prompting as well as detailed step-by-step prompting to guide the generation of the netlist.
We show the circuit and the best generated netlist by GPT-4o in Figure~\ref{fig:prompting-multimodal}.

Even in this best generation, the multimodal LLM makes several errors.
While it correctly identifies the number of transistors, it incorrectly classifies them.
It creates 3 PMOS and 8 NMOS transistors in the netlist; whereas the original circuit image has 7 PMOS and 4 NMOS transistors.
Secondly, there are issues in the actual connections as well, specially for nodes M3 and M4 where the Vin+ and Vin- are connected to the gates in the original circuit; while the multimodal model connects them to the drain.
Overall, the multimodal model have good image detection capabilities to identify the number of components and connections, but lack deeper recognition and reasoning capabilities to make correct circuit connections.

\begin{figure}
    \centering
    \includegraphics[width=0.98\linewidth]{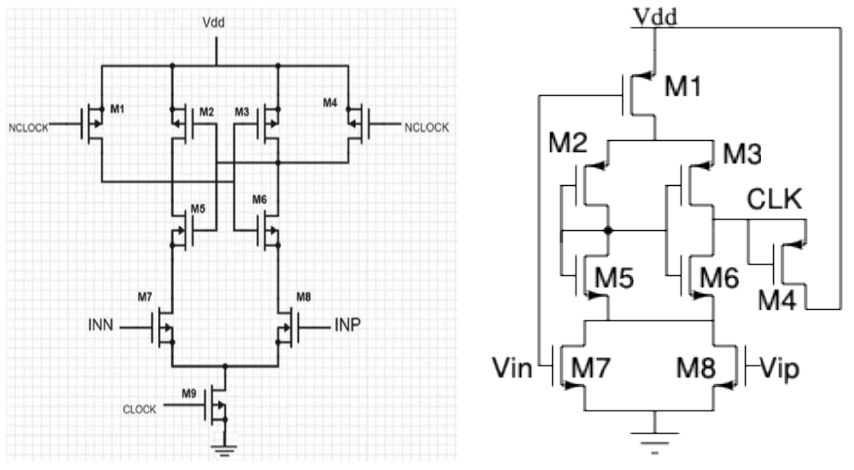}
    \Description{Highlighting the limitations and lack of generalization of utilizing existing fine-tuned models for netlist generation. (left) Original circuit, (right) Fine-tuned model generated incorrect circuit.}
    \caption{Highlighting the limitations and lack of generalization of utilizing existing fine-tuned models for netlist generation. (left) Original circuit, (right) Fine-tuned model generated incorrect circuit.}
    \label{fig:finetuning-prompting}
\end{figure}

\subsection{Prompting Fine-tuned models}

To overcome the issues of base multimodal LLMs, \cite{masala-chai} propose a fine-tuning recipe comprising a large dataset of simple schematic images and netlists.
They fine-tune a lightweight YOLO vision model \cite{yolo} specifically to detect non-standardized analog circuit symbols across varying image qualities.
They pass the detected components to an LLM to create the final netlist.
We evaluate their fine-tuned model + direct prompting of GPT-4o on 30 different test images of circuits related to SAR ADCs.
Simultaneously, we also fine-tune a GPT-4o model on >700 datapoints from \cite{amsnet} and evaluate on this test set.

Through our study, we conclude that both these fine-tuned models fail to generate correct netlists reliably.
We demonstrate an example of a failure case when generating a circuit image from \cite{sftimage} in Figure~\ref{fig:finetuning-prompting}, where we compare the original circuit (left) with the model-created circuit rendered into an image (right).
Firstly, the generated netlist completely removes one transistor M9.
Furthermore, the cross-coupled connections for M5-M6 and M2-M3 are incorrect.
The connections for M7-M8 are also shorted which is not the case in the original circuit
One of the major reasons we hypothesize for this poor performance is the distributional shift between the training and the testing data.
These models have been trained on simpler circuits from textbooks \cite{razavi, sedra} with high-quality images.
When tested on complex circuits with varying image quality, the model fails to generalize.
Largely, fine-tuned models also fail to incorporate extra rounds of text-based feedback owing to their training procedure; thus, it's difficult to correct the mistakes of these models without external models.

\section{Proposed Methodology - \modelName}
\label{sec:methodology}

To circumvent the limitations of open-ended unconstrained generation, we propose our multi-step LLM agentic framework \modelName.
Essentially, \modelName{} enhances the reliability and reduces LLM hallucinations by breaking down the complex analog design process into three smaller stages:
(1) Grounding in Expert Knowledge,
(2) Netlist Generation, and
(3) Netlist Sizing and Optimization.
For the scope of this work, we do not consider layout, as all our experiments are schematic-based.
For all the different agentic components we developed, a human expert tests them qualitatively to ensure reliability.
We provide an illustration of our agentic framework in Figure~\ref{fig:teaser} and provide a detailed overview below.

\subsection{Grounding in Expert Knowledge}
\label{sec:stage1-grounding}

One of the major flaws and limitations of using LLMs for analog design is hallucinations.
When the model is not as confident, it uses its best judgment and assumptions to fill the gaps.
While this is not as big an issue for other tasks, a wrong assumption at any stage of circuit design is highly costly and can have ripple effects, rendering the entire circuit incorrect.
We provided various illustrations of such hallucinations in Figure~\ref{fig:prompting-llms} and \ref{fig:prompting-multimodal}.

\begin{figure}
    \centering
    \includegraphics[width=0.8\linewidth]{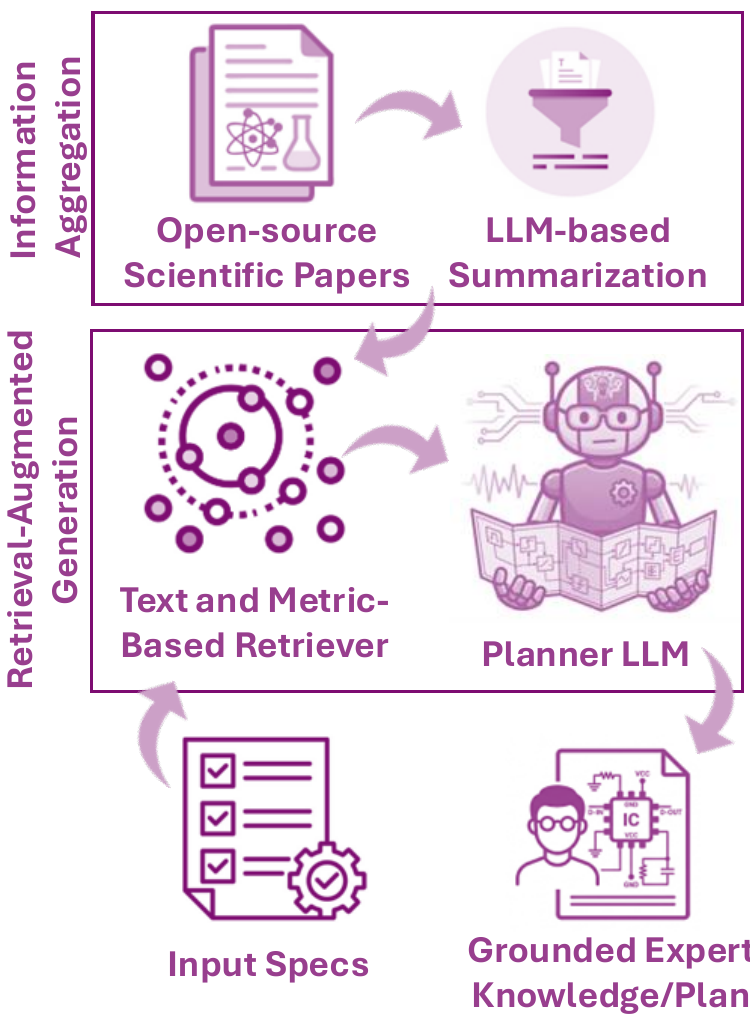}
    \Description{High-level illustration of stage 1 of our proposed \modelName{} involving information aggregation (top) from open-source scientific papers and a RAG-based Planner (bottom) for generating grounded plans for SAR ADC creation.}
    \caption{High-level illustration of stage 1 of our proposed \modelName{} involving information aggregation (top) from open-source scientific papers and a RAG-based Planner (bottom) for generating grounded plans for SAR ADC creation.}
    \label{fig:stage1-grounding}
\end{figure}

In order to reduce hallucinations, we propose to ground the LLM with expert analog design knowledge - specifically, scientific papers and findings.
We provide an illustrative workflow of this stage in Figure~\ref{fig:stage1-grounding}.
The workflow for this stage comprises two main steps: (1) Information Aggregation, and (2) Retrieval-Augmented Generation (RAG).
We describe each of these steps in more detail below.

\begin{figure*}
    \centering
    \includegraphics[width=0.95\linewidth]{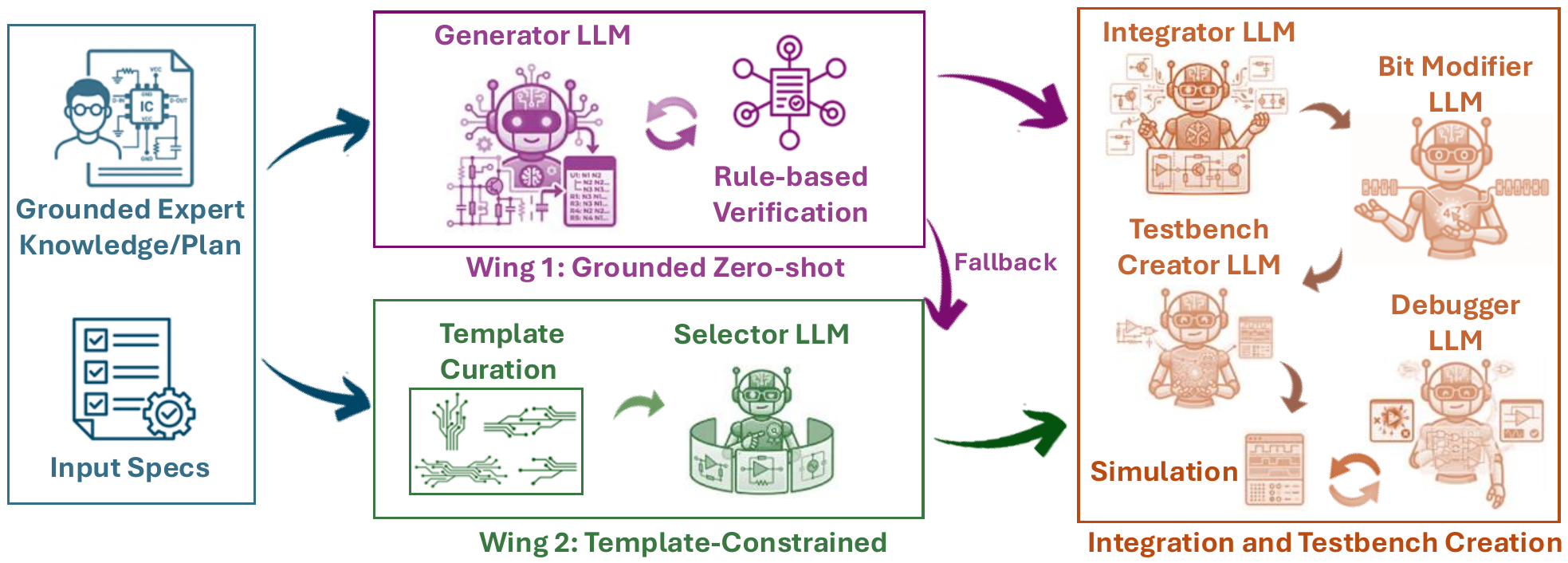}
    \Description{High-level illustration of stage 2 of our agentic framework \modelName{} which coordinates the creation of the ADC netlist based on the initial grounded expert knowledge and the user-specified input specs.}
    \caption{High-level illustration of stage 2 of our agentic framework \modelName{} which coordinates the creation of the ADC netlist based on the initial grounded expert knowledge and the user-specified input specs.}
    \label{fig:stage2-generation}
\end{figure*}

\subsubsection{Information Aggregation}
\label{sec:stage1-aggregation}

In order to ground the LLM, we first need to aggregate information from the analog design experts that we can share with the LLM.
Since we can not access and share proprietary company-specific information with LLMs, we rely on utilizing open-source data, specifically open-source scientific papers.
We start with the open-source survey and database from \cite{adc_survey}, comprising a large database of papers that have created scientific artifacts in the form of ADCs.
This database has information about the paper titles and links, as well as circuit-specific details like the technology node, signal-to-noise distortion ratio (SNDR), power, sampling frequency ($f_s$), etc.
Next, a human expert surveys and filters the papers and database entries related to SAR ADCs.
Finally, an LLM surveys the filtered papers and creates a small summary of the major findings for the circuits developed from each paper.
Overall, this aggregation of expert information serves as the main source of expert-grounded knowledge for the next steps in \modelName.

\subsubsection{Retrieval-Augmented Generation}
\label{sec:stage1-rag}

To utilize the aggregated information from the previous step, we utilize retrieval-augmented generation (RAG) with LLMs.
Specifically, based on the input description and specifications from the user, a heuristic retriever is used to match it to the best circuits from our database (based on proximity of metrics).
We then select the top $k$ circuits and provide its summary to the Planner LLM along with the target input specs and descriptions.
The Planner LLM then formulates a high-level plan for creating the SAR ADC and its individual components, grounded in the expert knowledge of the retrieved scientific papers.
This grounded expert knowledge/plan is passed to the next stage for the actual netlist generation.
We provide an illustration of the generated grounded plan in Figure~\ref{fig:plan-8-bit}.

\subsection{Netlist Generation}
\label{sec:stage2-generation}

Based on the grounded expert knowledge and plan for the SAR ADC generation from stage 1, the next and most important stage is to create the netlist.
At a high level, any netlist has several strongly intertwined components of the comparator, capdac, and the SAR logic - and our proposed workflow creates and combines them accordingly.
Overall, we have a two-winged netlist generation workflow:
(1) the first wing constructs the netlist through an iterative generation loop constrained by the expert literature from stage 1 and a rule-based verification engine, and
(2) the second wing utilizes template-constrained generation, where the LLM chooses and modifies one among the different templates based on the grounded expert data.
We present the high-level illustrative overview of this stage in Figure~\ref{fig:stage2-generation} and provide a detailed description below.

\subsubsection{Grounded Zero-shot Netlist Generation}
\label{sec:grounded-zs-generation}

We highlighted some limitations of LLM-based netlist generation in Section~\ref{sec:zero-shot-failure}.
To counter some of these limitations, we ground the LLM generations through expert data and rule-based feedback mechanisms.
Specifically, we prompt the Generator LLM with the grounded expert knowledge from stage 1, along with input specs, and component-specific information.
Based on this contextual information, the Generator is tasked with generating a netlist for the specified component from scratch.
Since LLMs lack niche analog design expertise, we create a rule-based verification engine that can study a netlist and provide feedback based on simple heuristics and rule violations.
Some of these rules include parameter matching across connections and singular-node warnings.
The Generator then updates the netlist based on the feedback provided from the rule-based verification engine, and the loop continues for a fixed number of iterations.

This wing of netlist generation is extremely liberal, as the LLM can add custom components and write logic accordingly.
At the same time, such unconstrained generation can amplify the possibility of errors.
We believe that this wing of netlist generation is futuristic and will improve netlist generation as the LLMs' inherent analog circuit-specific capabilities improve.
However, to reliably generate ADCs, we create a second fallback wing, which is triggered if we detect issues in the netlist from this first wing.
We describe this fallback generation method below.

\subsubsection{Template-constrained Netlist Generation}
\label{sec:template-constrained-generation}

Unconstrained generation through LLMs can lead to hallucinations and logical errors; thus, we provide an alternate way to generate ADCs via template-constrained generation.
To this end, we first curate a variety of different templates for the components of the SAR ADC, including the StrongArm latch \cite{comp1}, the Miyahara comparator \cite{comp2}, the standard binary-weighted capacitive DAC array \cite{capdac1, github-analog}, and the split-capacitor DAC array \cite{capdac2, maitreyi}.
Currently, we populate these templates using available open-source data and human effort; however, they can be easily populated with larger proprietary data as needed.
For each template, apart from the netlist, we utilize human experts / expert LLMs to curate a textual summary and the pros/cons.

Next, we move to the Selector LLM.
We provide the Selector with the templates for each component along with the grounded expert knowledge from Stage 1, and ask it to select the best template that would meet the input specs.
The meta-information about the templates proves crucial for the selection, especially for complex circuits where it is difficult for the LLM to directly understand the code.
Optionally, the LLM can also be prompted to update the netlist with minor modifications to the templates to cater to the input specs and grounded expert knowledge from stage 1.
Overall, this wing can be highly effective with a wide range of templates, but only provides little flexibility to customize the circuit to user inputs/needs.

\subsubsection{Integration and Testbench Creation}
\label{sec:testbench-creation}

Once the netlists for each comparator are created (via the direct or template-constrained wings), the final step remains to combine them and generate the higher-level logic.
Here, we prompt the Integrator LLM with the high-level parameters and the meta-information about the input and outputs of each component, as well as the original input specs and the grounded expert knowledge.
The Integrator is tasked to create the higher-level SAR ADC netlist which connects each of these components, adds additional gates/switches if needed, and provides the higher-level abstraction.
This step is relatively easier than direct netlist generation, as the LLM does not have to reason about the low-level SAR logic and transistor connections.

Since the templates/generated netlists can be for different number of bits relative to the input specs, the Bit-Modifier LLM updates the netlist to meet the specs.
Specifically, provided with the entire netlist and the target number of bits, the Modifier LLM updates the netlist across the different components to meet the target specs.

Next, we pass this netlist to the Testbench Creator LLM.
The Testbench Creator reads the higher-level SAR ADC netlist and creates a testbench to simulate the ADC.
Since unconstrained testbench creation can be noisy, we ground this LLM with a testbench template, which the LLM modifies based on the generated SAR ADC and the input specs.
Largely, the testbench utilizes an ideal DAC to convert the digital signals back to analog, and the evaluation is measured using the reconstruction quality.
Apart from this, a human expert also rectifies the testbench to validate the correctness of the reported metrics.

Finally, we loop this testbench with a simulation engine to execute the created netlist. 
If there are any errors/warnings as part of the simulation, this is passed to the Debugger LLM for modification and correction of the netlist.
This loop is run multiple times until the simulation succeeds to complete without any errors.

\begin{figure}
    \centering
    \includegraphics[width=\linewidth]{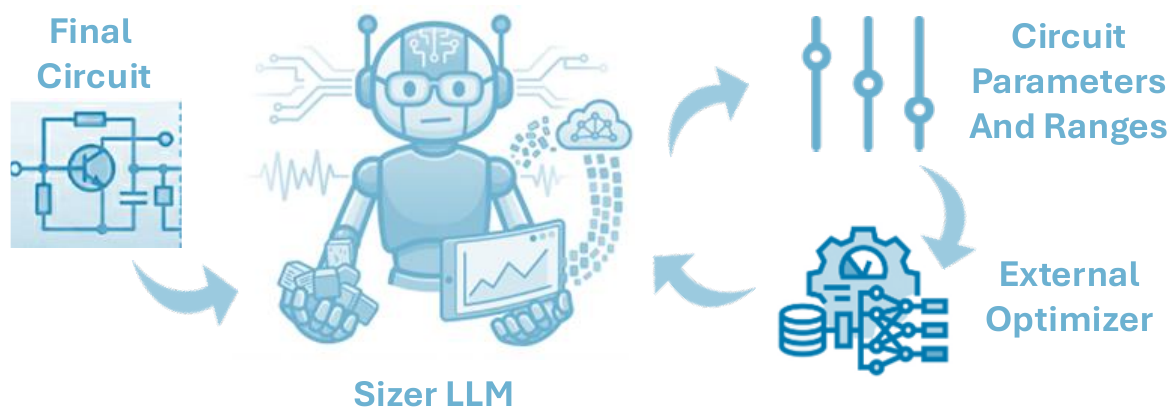}
    \Description{High-level illustration of stage 3 of our agentic framework \modelName{} - responsible for selecting and sizing the parameters of the final netlist.}
    \caption{High-level illustration of stage 3 of our agentic framework \modelName{} - responsible for selecting and sizing the parameters of the final netlist.}
    \label{fig:sizing-optimizing}
\end{figure}

\subsection{Netlist Sizing and Optimization}
\label{sec:stage3-sizing}

The final component after netlist generation remains sizing the different transistors and capacitors in the circuit.
Various recent works \cite{sizing1, sizing2, marl, eesizer} have studied sizing in-depth with and without LLMs and can be readily plugged in for this component.
In our work, we introduce this third stage of automated sizing using an LLM agent with an existing multi-objective optimizer, as illustrated in Figure~\ref{fig:sizing-optimizing}.
Specifically, given the final netlist from Stage 2, the Sizer LLM extracts the parameters that can be sized from the netlist based on the chosen templates.
These parameters have variable sharing where similar transistors/capacitors are assigned the same parameter variable.
Some set of parameters usually extracted by the Sizer include transistor-level and switch-level length, width, multiplicity ($m$), number of fingers ($nf$), as well as unit capacitance ($cc$) for capacitors. 

Based on the selected parameters, the Sizer writes a full-loop of optimization script, where it supplies these parameters and their expected ranges (from prior expert knowledge) to an external multi-objective optimizer, which is then run across a wide range of iterations.
After the optimization run, the results and logs are presented to the Sizer, which then updates the parameter set along with the range of search - an updated version of a previous work LEDRO \cite{ledro}.
We conduct this iterative loop a few times and the final sized netlist constitutes the final SAR ADC developed by the LLM.

\section{Synthesis of 8-bit SAR ADC}
\label{sec:main-synthesis}

Here, we demonstrate the efficacy of \modelName{} by developing a low-power 8-bit SAR ADC end-to-end using our framework.
First, we provide the experimental setup and some details about the implementation details.
Later, we present our main simulated results of the synthesized SAR ADC from our framework.

\begin{table}[t]
    \centering
    \begin{tabular}{l|ccccc}
        \toprule
        & \textbf{ENOB} & \textbf{SINAD} & \textbf{SFDR} & \textbf{THD} & \textbf{Power} \\
        \midrule
        Target & 7.5 & 45db & 50db & -50db & 8uW \\
        \midrule
        Achieved & 7.59 & 47.5db & 55.6db & -51.5db & 6uW \\
        \bottomrule
    \end{tabular}
    \caption{Main results comparing \modelName's achieved specs against the target specs for the low-power 8-bit SAR ADC on the Cadence GPDK-45nm technology node.}
    \label{tab:main-results}
\end{table}

\subsection{Experimental Setup}
Our major experiments are conducted on the Cadence GPDK-45nm technology node, simulated through Cadence Spectre.
To be fair with our studies of limitation analysis in Section~\ref{sec:zero-shot-failure}, we utilize GPT-4o as the main LLM for our experiments.
Our input specifications comprise the number of input bits (\# bits), sampling frequency ($f_s$), Effective Number of Bits (ENOB), Signal-to-Noise and Distortion (SINAD), Spurious-Free Dynamic Range (SFDR), Total Harmonic Distortion (THD), power, as well as an optional text description of the SAR ADC for human-in-the-loop guidance/steering.
For power calculations, we average the DAC, comparator, and the SAR logic powers.

For our retriever in stage 1, we utilize a simple k-Nearest Neighbors (k-NN) retrieval based on Euclidean distance in a min-max normalized feature space using $k=5$.
This ensures that our selected papers for grounding the LLMs are similar to the target user specs.
For our template-constrained netlist generation in stage 2, we create templates for three components of the SAR ADC pipeline - specifically the comparator, the SAR logic, and the capacitor DAC array (capdac).
These components are populated with the open-source and human expert curated templates as detailed in Section~\ref{sec:template-constrained-generation}.
For our testbench template, we set VDD to 1 or 1.2 V depending on specification.
For the optimizer in stage 3, we use multi-objective Bayesian optimization featuring Sobol-initialized designs, independent Gaussian-process surrogates (BoTorch) \cite{botorch}, and qParEGO-style batch expected improvement under random Chebyshev scalarizations with QMC \cite{parego}.

\subsection{SAR ADC Synthesis Results}

For our primary demonstration, we consider developing a low-power 8-bit SAR ADC with human-expert provided target specifications (specs), as shown in Table~\ref{tab:main-results} (first row).
We set the sampling frequency ($f_s$) to 3.33 MHz and clock frequency ($f_{clk}$) as 20 MHz.
We provide the input text description simply as ``synchronous low power".
We present the achieved metrics of our generated 8-bit SAR ADC against the target user specs in Table~\ref{tab:main-results}.
As noted, the \modelName's generated SAR ADC meets all the user specifications comfortably.

\begin{figure}
    \centering
    \includegraphics[width=0.95\linewidth]{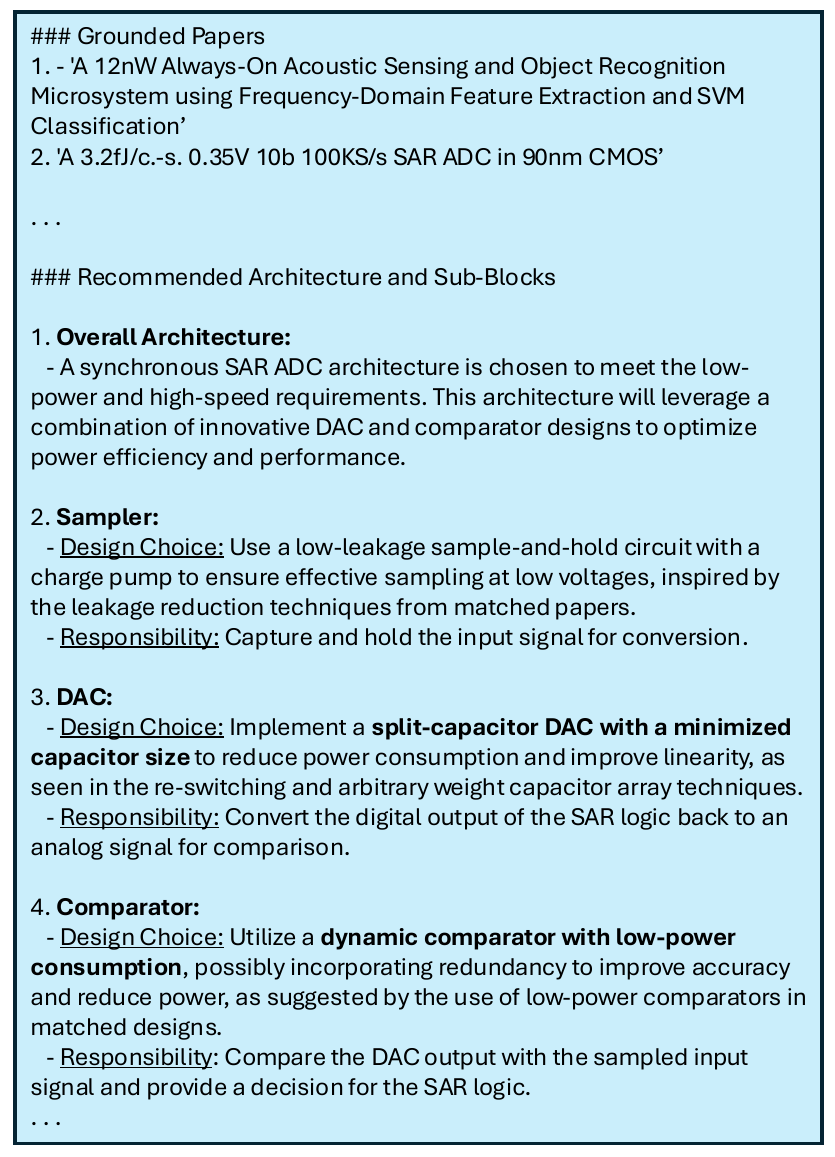}
    \Description{Illustration of the grounded papers and the detailed plan generated by the Planner LLM for the 8-bit SAR ADC. We highlight some of the key choices it recommends.}
    \caption{Illustration of the grounded papers and the detailed plan generated by the Planner LLM for the 8-bit SAR ADC. We highlight some of the key choices it recommends.}
    \label{fig:plan-8-bit}
\end{figure}

\begin{figure}
    \centering
    \includegraphics[width=0.95\linewidth]{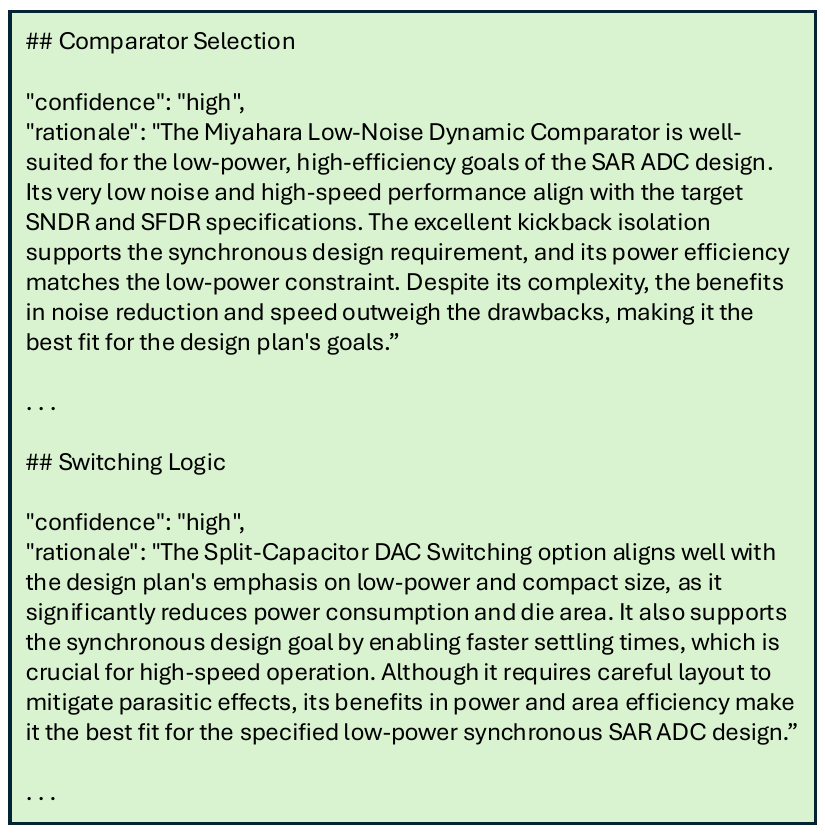}
    \Description{Illustrating the choices and the underlying rationale for template selection by our Selection LLM for the generation of the low-power 8-bit SAR ADC.}
    \caption{Illustrating the choices and the underlying rationale for template selection by our Selection LLM for the generation of the low-power 8-bit SAR ADC.}
    \label{fig:selection-8-bit}
\end{figure}

To make a deeper dive into our multi-agent framework's decision-making, we study its intermediate outputs.
In Figure~\ref{fig:plan-8-bit}, we show some of the retrieved grounded papers and part of the detailed grounded plan generated by the Planner in stage 1, highlighting some of its design choices.
The grounded zero-shot netlist generation had some errors and didn't succeed, but the fallback to template-constrained netlist succeeded.
Some of the choices are well-grounded and we illustrate the Selector's rationale in Figure~\ref{fig:selection-8-bit}.

\section{Generalization Analysis}
\label{sec:generalization}

To demonstrate the generalizability of \modelName{} to generate broader range of SAR ADCs, we conduct various additional analytical experiments - which we detail more in this section.

\begin{table}[t]
    \centering
    \begin{tabular}{l|ccccc}
        \toprule
        & \textbf{ENOB} & \textbf{SINAD} & \textbf{SFDR} & \textbf{THD} & \textbf{Power} \\
        \midrule
        Target & 7.5 & 45dB & 50dB & -50dB & 8$\mu$W \\
        \midrule
        Achieved & 7.82 & 48.8dB & 56.5dB & -52.5dB & 5.7$\mu$W \\
        \bottomrule
    \end{tabular}
    \caption{Results highlighting the generalization across technology nodes by comparing the achieved specs against the target specs for the 8-bit SAR ADC on a different technology node of TSMC-65nm.}
    \label{tab:tsmc-results}
\end{table}

\subsection{Technology Nodes}

Here, we show the generalization of \modelName's SAR ADC across technology nodes - specifically from GPDK-45nm to TSMC-65nm node.
Owing to privacy and copyright concerns with sharing foundry's technology node files, directly running our framework to generate the SAR ADC is unfeasible (although we believe doing so will further improve the ADC).
Instead, the LLM-generated 8-bit SAR ADC was migrated from Cadence GPDK-45nm to TSMC-65nm using custom scripts developed to automatically map and substitute the technology primitives.
We utilize a local optimizer for sizing instead of our LLM framework to ensure no foundry data leakage.
Despite no technology node-specific optimizations, the generated SAR ADC performs well out-of-the-box with the main metrics compared with the target user specs provided in Table~\ref{tab:tsmc-results}.
Overall, this demonstrates the transferability and generalization of \modelName{} across technology nodes.

\begin{table}[t]
    \centering
    \begin{tabular}{l|ccccc}
        \toprule
        & \textbf{ENOB} & \textbf{SINAD} & \textbf{SFDR} & \textbf{THD} & \textbf{Power} \\
        \midrule
        Target & 3.7 & 24dB & 31.5dB & -28dB & 25$\mu$W \\
        \midrule
        Achieved & 3.86 & 25dB & 33.6dB & -29.8dB & 21.1$\mu$W \\
        \bottomrule
    \end{tabular}
    \caption{Results highlighting the bit modification provided input specifications by comparing the achieved specs against the target specs for an asynchronous 4-bit SAR ADC design.}
    \label{tab:4-bit-results}
\end{table}

\subsection{Input Specifications}

Next, we conduct experiments to evaluate the adaptability of \modelName{} to build customized SAR ADCs for varying user input specifications.
Specifically, we update the new user target specs and provide the metrics in Table~\ref{tab:4-bit-results} (top row).
We set the sampling frequency ($f_s$) to 20 MHz, choose asynchronous design, and the number of bits to 4.
For the 4-bit SAR ADC, the Selector LLM chose simpler templates compared to the previous 8-bit design.
It selected a standard bottom-plate capdac instead of a split-capacitor array, as the lower resolution does not require capacitance scaling.
Additionally, it replaced the low-noise Miyahara comparator with a standard StrongArm latch, since kickback noise is far less critical at 4 bits and gives higher speed.
We present the performance metrics of the ADC developed by our framework in Table~\ref{tab:4-bit-results} (bottom row).
The results show how it meets the specs.
In fact, \modelName{} removed a transistor connected in parallel in the bootstrap switch in the template with the rationale of reducing THD and that indeed improved performance.
Overall, this analysis provides a proof-of-concept for the generalizability of \modelName{} across different input specifications.



\begin{table}[t]
    \centering
    \begin{tabular}{l|ccccc}
        \toprule
        & \textbf{ENOB} & \textbf{SINAD} & \textbf{SFDR} & \textbf{THD} & \textbf{Power} \\
        \midrule
        Target & 9.5 & 58dB & 68dB & -65dB & 7.5$\mu$w \\
        \midrule
        Achieved & 8.93 & 55.5dB & 62.9dB & -59.9dB & 6.1$\mu$W \\
        \bottomrule
    \end{tabular}
    \caption{Results highlighting the bit modification capability of our agentic framework \modelName{} when asked to modify the 8-bit SAR ADC for 10 bits.}
    \label{tab:10-bit-results}
\end{table}

\begin{figure}
    \centering
    \includegraphics[width=0.98\linewidth]{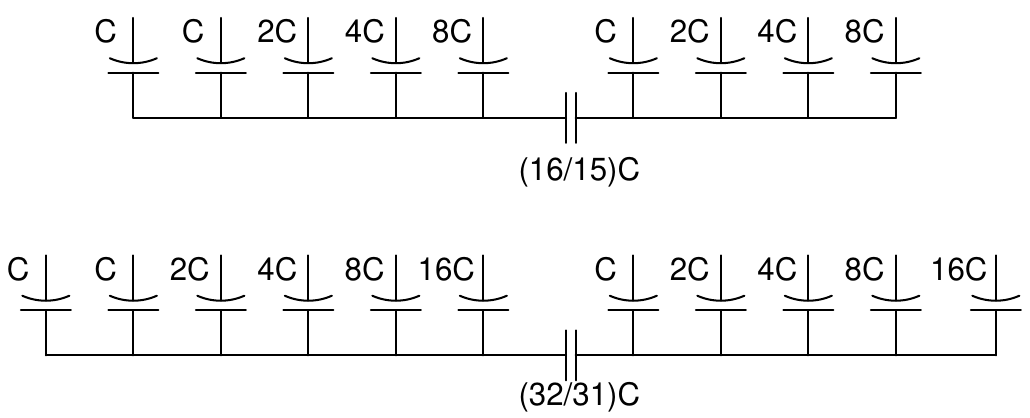}
    \Description{Illustration highlighting the bit-modification capability of our framework where \modelName{} extended a 8-bit capdac array (top) to 10 bits (bottom).}
    \caption{Illustration highlighting the bit-modification capability of our framework where \modelName{} extended a 8-bit capdac array (top) to 10 bits (bottom).}
    \label{fig:capdac-extension}
\end{figure}

\subsection{Bit Modification}

In this analysis, we study the capability of the framework to adapt its circuit for varying number of bits.
Specifically, this is a test of the Bit Modifier Agent of \modelName{} to modify the template-based circuit (built for a specific number of bits) to match the target user specs.
For this analysis, we utilize Gemini 3.1 as the LLM for the Bit Modifier.
To test this, we use the same user specifications as the main synthesis experiment for the 8-bit SAR ADC, but only change the number of bits to 10 and set the sampling frequency ($f_s$) as 2.5 MHz.
Noting here that the templates were originally created specifically for 8 bits, and their performance when extrapolated to 10 bits remains unknown/sub-optimal.
We report the results for the 10-bit SAR ADC created by \modelName{} using the Bit Modifier's capability in Table~\ref{tab:10-bit-results}.
We observe that the ADC fails to meet the target specs, but achieves close to 9 ENOB.
Upon human expert inspection, we conclude that the Bit Modifier has strong pattern-matching reasoning to extrapolate the circuit, as we show an illustration of the correct extension of the capdac array in Figure~\ref{fig:capdac-extension}.


\section{Conclusion and Future Work}
\label{sec:conclusion}

In this paper, we propose a template-constrained LLM agentic framework \modelName{} that successfully bridges the gap between unconstrained text generation and the rigorous physical demands of analog EDA.
By grounding the LLM's generative search space in expert-defined architectural priors, our framework serves as a proof-of-concept for the end-to-end synthesis of functional SAR ADCs across multiple specifications and technology nodes.
While the current scope of this work validates the methodology on basic SAR ADC topologies, it establishes a reliable, non-hallucinating paradigm for integrating LLMs into EDA workflows.
Future work can focus on expanding the granularity and diversity of the template libraries, as well as leveraging the evolving reasoning capabilities of newer LLMs, to tackle the continuous-domain optimization of much more complex mixed-signal systems.
Ultimately, we believe this structured, prior-grounded approach opens a highly promising avenue for reliable AI-assisted analog circuit design.

\bibliography{ref}

@inproceedings{comp1,
  title={A current-mode latch sense amplifier and a static power saving input buffer for low-power architecture},
  author={Kobayashi, Tsuguo and Nogami, Kazutaka and Shirotori, Tsukasa and Fujimoto, Yukihiro and Watanabe, Osamu},
  booktitle={1992 symposium on VLSI circuits digest of technical papers},
  pages={28--29},
  year={1992},
  organization={IEEE}
}

@inproceedings{comp2,
  title={A low-noise self-calibrating dynamic comparator for high-speed ADCs},
  author={Miyahara, Masaya and Asada, Yusuke and Paik, Daehwa and Matsuzawa, Akira},
  booktitle={2008 IEEE Asian Solid-State Circuits Conference},
  pages={269--272},
  year={2008},
  organization={IEEE}
}

@misc{adc_survey,
   author = {Murmann, Boris},
   title = {{ADC Performance Survey 1997-2026}},
   note = {[Online]. Available: \url{https://github.com/bmurmann/ADC-survey}}
}

@article{capdac1,
  title={All-MOS charge redistribution analog-to-digital conversion techniques. I},
  author={McCreary, James L and Gray, Paul R},
  journal={IEEE Journal of Solid-State Circuits},
  volume={10},
  number={6},
  pages={371--379},
  year={1975},
  publisher={IEEE}
}

@article{capdac2,
  title={A two-stage weighted capacitor network for D/AA/D conversion},
  author={Yee, YS and Terman, LM and Heller, LG},
  journal={IEEE Journal of Solid-State Circuits},
  volume={14},
  number={4},
  pages={778--781},
  year={2003},
  publisher={IEEE}
}

@misc{masala-chai,
      title={Masala-CHAI: A Large-Scale SPICE Netlist Dataset for Analog Circuits by Harnessing AI}, 
      author={Bhandari, Jitendra and Bhat, Vineet and He, Yuheng and Rahmani, Hamed and Garg, Siddharth and Karri, Ramesh},
      year={2025},
      eprint={2411.14299},
      archivePrefix={arXiv},
      primaryClass={cs.AR},
      url={https://arxiv.org/abs/2411.14299}, 
}

@INPROCEEDINGS{low-offset,
  author={Sachdeva, Yogesh and Nehra, Nalin and Bansal, Shikhar and Garima},
  booktitle={2021 Second International Conference on Electronics and Sustainable Communication Systems (ICESC)}, 
  title={Review of Dynamic Comparators for ADCs}, 
  year={2021},
  volume={},
  number={},
  pages={83-90},
  doi={10.1109/ICESC51422.2021.9532865}}

@inproceedings{dtcomp,
  title={A double-tail latch-type voltage sense amplifier with 18ps setup+ hold time},
  author={Schinkel, Daniel and Mensink, Eisse and Klumperink, Eric and Van Tuijl, Ed and Nauta, Bram},
  booktitle={2007 IEEE international solid-state circuits conference. Digest of technical papers},
  pages={314--605},
  year={2007},
  organization={IEEE}
}

@incollection{sftimage,
  title={Design and comparative analysis of dynamic comparators for SAR ADC},
  author={Ansari, Noman Ahmed and Jaiswal, Priyansh and Tyagi, Mohit and Mittal, Poornima},
  booktitle={Emerging Electronics and Automation: Select Proceedings of E2A 2021},
  pages={391--401},
  year={2022},
  publisher={Springer}
}

@inproceedings{ledro,
  title={Ledro: Llm-enhanced design space reduction and optimization for analog circuits},
  author={Kochar, Dimple Vijay and Wang, Hanrui and Chandrakasan, Anantha P and Zhang, Xin},
  booktitle={2025 IEEE International Conference on LLM-Aided Design (ICLAD)},
  pages={141--148},
  year={2025},
  organization={IEEE}
}

@article{botorch,
  title={BoTorch: A framework for efficient Monte-Carlo Bayesian optimization},
  author={Balandat, Maximilian and Karrer, Brian and Jiang, Daniel and Daulton, Samuel and Letham, Ben and Wilson, Andrew G and Bakshy, Eytan},
  journal={Advances in neural information processing systems},
  volume={33},
  pages={21524--21538},
  year={2020}
}

@article{parego,
  title={Differentiable expected hypervolume improvement for parallel multi-objective Bayesian optimization},
  author={Daulton, Samuel and Balandat, Maximilian and Bakshy, Eytan},
  journal={Advances in neural information processing systems},
  volume={33},
  pages={9851--9864},
  year={2020}
}

@article{analogseeker,
  title={AnalogSeeker: An Open-source Foundation Language Model for Analog Circuit Design},
  author={Chen, Zihao and Zhuang, Ji and Shen, Jinyi and Ke, Xiaoyue and Yang, Xinyi and Zhou, Mingjie and Du, Zhuoyao and Yan, Xu and Wu, Zhouyang and Xu, Zhenyu and others},
  journal={arXiv preprint arXiv:2508.10409},
  year={2025}
}

@article{lamagic,
  title={Lamagic: Language-model-based topology generation for analog integrated circuits},
  author={Chang, Chen-Chia and Shen, Yikang and Fan, Shaoze and Li, Jing and Zhang, Shun and Cao, Ningyuan and Chen, Yiran and Zhang, Xin},
  journal={arXiv preprint arXiv:2407.18269},
  year={2024}
}

@inproceedings{artisan,
  title={Artisan: Automated operational amplifier design via domain-specific large language model},
  author={Chen, Zihao and Huang, Jiangli and Liu, Yiting and Yang, Fan and Shang, Li and Zhou, Dian and Zeng, Xuan},
  booktitle={Proceedings of the 61st ACM/IEEE Design Automation Conference},
  pages={1--6},
  year={2024}
}

@article{ampagent,
  title={Ampagent: An llm-based multi-agent system for multi-stage amplifier schematic design from literature for process and performance porting},
  author={Liu, Chengjie and Chen, Weiyu and Peng, Anlan and Du, Yuan and Du, Li and Yang, Jun},
  journal={arXiv preprint arXiv:2409.14739},
  year={2024}
}

@inproceedings{analogxpert,
  title={Analogxpert: Automating analog topology synthesis by incorporating circuit design expertise into large language models},
  author={Zhang, Haoyi and Sun, Shizhao and Lin, Yibo and Wang, Runsheng and Bian, Jiang},
  booktitle={2025 International Symposium of Electronics Design Automation (ISEDA)},
  pages={772--777},
  year={2025},
  organization={IEEE}
}

@article{chipnemo,
  title={Chipnemo: Domain-adapted llms for chip design},
  author={Liu, Mingjie and Ene, Teodor-Dumitru and Kirby, Robert and Cheng, Chris and Pinckney, Nathaniel and Liang, Rongjian and Alben, Jonah and Anand, Himyanshu and Banerjee, Sanmitra and Bayraktaroglu, Ismet and others},
  journal={arXiv preprint arXiv:2311.00176},
  year={2023}
}

@article{atelier,
  title={Atelier: An automated analog circuit design framework via multiple large language model-based agents},
  author={Shen, Jinyi and Chen, Zihao and Zhuang, Ji and Huang, Jiangli and Yang, Fan and Shang, Li and Bi, Zhaori and Yan, Changhao and Zhou, Dian and Zeng, Xuan},
  journal={IEEE Transactions on Computer-Aided Design of Integrated Circuits and Systems},
  year={2025},
  publisher={IEEE}
}

@inproceedings{analogcoder,
  title={Analogcoder: Analog circuit design via training-free code generation},
  author={Lai, Yao and Lee, Sungyoung and Chen, Guojin and Poddar, Souradip and Hu, Mengkang and Pan, David Z and Luo, Ping},
  booktitle={Proceedings of the AAAI Conference on Artificial Intelligence},
  volume={39},
  number={1},
  pages={379--387},
  year={2025}
}

@article{maitreyi,
  title={Randomized switching SAR (RS-SAR) ADC for power and EM side-channel security},
  author={Ashok, Maitreyi and Levine, Edlyn V and Chandrakasan, Anantha P},
  journal={IEEE Solid-State Circuits Letters},
  volume={5},
  pages={247--250},
  year={2022},
  publisher={IEEE}
}

@misc{github-analog,
  author       = {Muhammad Aldacher},
  title        = {Analog Design of Asynchronous SAR ADC},
  year         = {2020},
  publisher    = {GitHub},
  journal      = {GitHub repository},
  howpublished = {\url{https://github.com/muhammadaldacher/Analog-Design-of-Asynchronous-SAR-ADC}}
}

@article{ladac,
  title={Ladac: Large language model-driven auto-designer for analog circuits},
  author={Liu, Chengjie and Liu, Yijiang and Du, Yuan and Du, Li},
  journal={Authorea Preprints},
  year={2024},
  publisher={Authorea}
}

@inproceedings{amsnet,
  title={Amsnet: Netlist dataset for ams circuits},
  author={Tao, Zhuofu and Shi, Yichen and Huo, Yiru and Ye, Rui and Li, Zonghang and Huang, Li and Wu, Chen and Bai, Na and Yu, Zhiping and Lin, Ting-Jung and others},
  booktitle={2024 IEEE LLM Aided Design Workshop (LAD)},
  pages={1--5},
  year={2024},
  organization={IEEE}
}

@article{sizing1,
  title={A Global-Local Optimization Approach for Asynchronous SAR ADC Design},
  author={Hao, Yijia and Li, Ken and Gandara, Miguel and Li, Shaolan and Liu, Bo},
  journal={arXiv preprint arXiv:2507.19541},
  year={2025}
}

@inproceedings{sizing2,
  title={Automated SAR ADC Sizing Using Analytical Equations},
  author={Li, Zhongyi and Tao, Zhuofu and Zhou, Yanze and Shi, Yichen and Yu, Zhiping and Lin, Ting-Jung and He, Lei},
  booktitle={2025 International Symposium of Electronics Design Automation (ISEDA)},
  pages={214--220},
  year={2025},
  organization={IEEE}
}

@article{marl,
  title={Multiagent based reinforcement learning (MA-RL): An automated designer for complex analog circuits},
  author={Bao, Jiarui and Zhang, Jinxin and Huang, Zhangcheng and Bi, Zhaori and Feng, Xingwei and Zeng, Xuan and Lu, Ye},
  journal={IEEE Transactions on Computer-Aided Design of Integrated Circuits and Systems},
  volume={43},
  number={12},
  pages={4398--4411},
  year={2024},
  publisher={IEEE}
}

@article{eesizer,
  title={Eesizer: Llm-based ai agent for sizing of analog and mixed signal circuit},
  author={Liu, Chang and Chitnis, Danial},
  journal={IEEE Transactions on Circuits and Systems I: Regular Papers},
  year={2025},
  publisher={IEEE}
}

@article{heart,
  title={HeaRT: A Hierarchical Circuit Reasoning Tree-Based Agentic Framework for AMS Design Optimization},
  author={Poddar, Souradip and Ho, Chia-Tung and Wei, Ziming and Cao, Weidong and Ren, Haoxing and Pan, David Z},
  journal={arXiv preprint arXiv:2511.19669},
  year={2025}
}

@article{ref1,
  title={AI-Driven Integrated Circuit Design: A Survey of Techniques, Challenges, and Opportunities},
  author={Guven, Islam and Parlak, Mehmet and Lederer, Dimitri and De Vleeschouwer, Christophe},
  journal={IEEE Access},
  year={2025},
  publisher={IEEE}
}

@article{ref2,
  title={The dawn of ai-native eda: Opportunities and challenges of large circuit models},
  author={Chen, Lei and Chen, Yiqi and Chu, Zhufei and Fang, Wenji and Ho, Tsung-Yi and Huang, Ru and Huang, Yu and Khan, Sadaf and Li, Min and Li, Xingquan and others},
  journal={arXiv preprint arXiv:2403.07257},
  year={2024}
}

@article{grposmu,
  title={GRPO with State Mutations: Improving LLM-Based Hardware Test Plan Generation},
  author={Kochar, Dimple Vijay and Pinckney, Nathaniel and Liu, Guan-Ting and Ho, Chia-Tung and Deng, Chenhui and Ren, Haoxing and Khailany, Brucek},
  journal={arXiv preprint arXiv:2601.07593},
  year={2026}
}

@article{swe,
  title={Swe-bench: Can language models resolve real-world github issues?},
  author={Jimenez, Carlos E and Yang, John and Wettig, Alexander and Yao, Shunyu and Pei, Kexin and Press, Ofir and Narasimhan, Karthik},
  journal={arXiv preprint arXiv:2310.06770},
  year={2023}
}

@inproceedings{opensar1,
  title={OpenSAR: An open source automated end-to-end SAR ADC compiler},
  author={Liu, Mingjie and Tang, Xiyuan and Zhu, Keren and Chen, Hao and Sun, Nan and Pan, David Z},
  booktitle={2021 IEEE/ACM International Conference On Computer Aided Design (ICCAD)},
  pages={1--9},
  year={2021},
  organization={IEEE}
}

@article{opensar2,
  title={1-and 80-MS/s SAR ADCs in 40-nm CMOS with end-to-end compilation},
  author={Liu, Mingjie and Tang, Xiyuan and Zhu, Keren and Chen, Hao and Sun, Nan and Pan, David Z},
  journal={IEEE Solid-State Circuits Letters},
  volume={5},
  pages={292--295},
  year={2022},
  publisher={IEEE}
}

@article{ding2018hybrid,
  title={A hybrid design automation tool for SAR ADCs in IoT},
  author={Ding, Ming and Harpe, Pieter and Chen, Guibin and Busze, Benjamin and Liu, Yao-Hong and Bachmann, Christian and Philips, Kathleen and Van Roermund, Arthur},
  journal={IEEE Transactions on Very Large Scale Integration (VLSI) Systems},
  volume={26},
  number={12},
  pages={2853--2862},
  year={2018},
  publisher={IEEE}
}

@article{seo2018reusable,
  title={A reusable code-based SAR ADC design with CDAC compiler and synthesizable analog building blocks},
  author={Seo, Min-Jae and Roh, Yi-Ju and Chang, Dong-Jin and Kim, Wan and Kim, Ye-Dam and Ryu, Seung-Tak},
  journal={IEEE Transactions on circuits and systems II: Express Briefs},
  volume={65},
  number={12},
  pages={1904--1908},
  year={2018},
  publisher={IEEE}
}

@article{settaluri2020autockt,
  title={Autockt: Deep reinforcement learning of analog circuit designs},
  author={Settaluri, Keertana and Haj-Ali, Ameer and Huang, Qijing and Hakhamaneshi, Kourosh and Nikolic, Borivoje},
  journal={arXiv preprint arXiv:2001.01808},
  year={2020}
}

@inproceedings{wang2020gcnrl,
  title={GCN-RL circuit designer: Transferable transistor sizing with graph neural networks and reinforcement learning},
  author={Wang, Hanrui and Wang, Kuan and Yang, Jiacheng and Shen, Linxiao and Sun, Nan and Lee, Hae-Seung and Han, Song},
  booktitle={2020 57th ACM/IEEE Design Automation Conference (DAC)},
  pages={1--6},
  year={2020},
  organization={IEEE}
}

@inproceedings{zhang2020efficient,
  title={An efficient asynchronous batch Bayesian optimization approach for analog circuit synthesis},
  author={Zhang, Shuhan and Yang, Fan and Zhou, Dian and Zeng, Xuan},
  booktitle={2020 57th ACM/IEEE design automation conference (DAC)},
  pages={1--6},
  year={2020},
  organization={IEEE}
}

@article{mandal2002cmos,
  title={CMOS op-amp sizing using a geometric programming formulation},
  author={Mandal, Pradip and Visvanathan, V},
  journal={IEEE Transactions on Computer-Aided Design of Integrated circuits and systems},
  volume={20},
  number={1},
  pages={22--38},
  year={2002},
  publisher={IEEE}
}

@article{gpt4o,
  title={Gpt-4o system card},
  author={Hurst, Aaron and Lerer, Adam and Goucher, Adam P and Perelman, Adam and Ramesh, Aditya and Clark, Aidan and Ostrow, AJ and Welihinda, Akila and Hayes, Alan and Radford, Alec and others},
  journal={arXiv preprint arXiv:2410.21276},
  year={2024}
}

@article{deepseek,
  title={Deepseek-v3 technical report},
  author={Liu, Aixin and Feng, Bei and Xue, Bing and Wang, Bingxuan and Wu, Bochao and Lu, Chengda and Zhao, Chenggang and Deng, Chengqi and Zhang, Chenyu and Ruan, Chong and others},
  journal={arXiv preprint arXiv:2412.19437},
  year={2024}
}

@misc{gemini,
  author       = {{Google DeepMind}},
  title        = {Gemini 3 Pro Model Card},
  year         = {2026},
  howpublished = {\url{https://storage.googleapis.com/deepmind-media/Model-Cards/Gemini-3-Pro-Model-Card.pdf}},
  publisher    = {Google},
  note         = {Accessed: 2026-04-12}
}

@article{yin2024survey,
  title={A survey on multimodal large language models},
  author={Yin, Shukang and Fu, Chaoyou and Zhao, Sirui and Li, Ke and Sun, Xing and Xu, Tong and Chen, Enhong},
  journal={National Science Review},
  volume={11},
  number={12},
  pages={nwae403},
  year={2024},
  publisher={Oxford University Press}
}

@book{razavi,
author = {Razavi, Behzad},
title = {Design of Analog CMOS Integrated Circuits},
year = {2000},
isbn = {0072380322},
publisher = {McGraw-Hill, Inc.},
address = {USA},
}

@misc{sedra,
  title={Microelectronic Circuits pp. 263-267},
  author={Sedra, Adel S and Smith, Kenneth C},
  year={1987},
  publisher={Holt, Rinehart and Winston Inc., New York}
}

@misc{yolo,
  author = {Jocher, Glenn and Chaurasia, Ayush and Qiu, Jing},
  title = {Ultralytics YOLO},
  version = {8.0.0},
  year = {2023},
  url = {https://github.com/ultralytics/ultralytics},
  note = {Computer software}
}

@String{Computer = "{IEEE} Computer" }

@String{AMS = "American Mathematical Society" }

@String{Springer = "Springer-Verlag" }

\end{document}